\documentstyle[12pt,aaspp4]{article}
\def\tempest%
{\begin{array}{ccc}
1 & 1 & 1 \\
1 & 1 & 1 \\
4 & 3 & 8
\end{array}}
\def\kms{{\rm km}\,{\rm s}^{-1}}
\def\kpc{{\rm kpc}}

\def\min{{\rm min}}

\def\lim{{\rm lim}}

\begin{document}

\title{Analytic Study of Grid Star and Reference Star Selection for
the {\it Space Interferometry Mission}}

\author{
Andrew Gould
}
\affil{Ohio State University, Department of Astronomy, 
140 W.\ 18th Ave., Columbus, OH 43210, USA; gould@astronomy.ohio-state.edu} 
\centerline{and}
\affil{Laboratoire de Physique Corpusculaire et Cosmologie,
College de France, 11 pl.\ Marcelin Berthelot, F-75231, Paris, France}


\begin{abstract}

Grid stars and reference stars provide the fundamental global and
local astrometric reference frames for observations by the 
{\it Space Interferometry Mission}.  They must therefore
be astrometrically stable at the $\sim 1\,\mu$as level.  I present
simple formulae in closed form to estimate contamination of these 
frames by motions due to stellar companions that go undetected in 
a radial-velocity (RV) survey of specified precision.  The 
contamination rate depends almost entirely on the binary-period 
distribution function and not on the details of the mass or 
eccentricity distributions.  Screening by a modest RV survey 
($\sigma_{\rm rv}=60\,\rm m\,s^{-1}$) can reduce the fraction of grid
stars having detectable unmodeled accelerations to $\ll 1\%$.
Reference-star selection promises to be much more challenging, partly
because the requirements of astrometric stability are more severe
and partly because the required density of sources is $\sim 100$
times higher, which latter implies that less satisfactory candidates 
will have to be accepted.  The tools presented here can help design 
a reference-star selection strategy, but a full treatment of the 
problem will require better knowledge of their companions in the 
planetary mass range.

\end{abstract}

\vfil\eject 
 
\section{Introduction}

	By their nature, astrometric observations measure the angular
separation of one star relative to another, either absolutely or 
projected on some axis.  Hence, if attention is restricted only to
these two stars, one cannot determine which, if either, is stationary,
and which is moving.  

	There are basically three approaches to resolving this ambiguity.
The traditional method is to find reference stars that, based on their
photometric properties, are believed to be so far away that their
astrometric motions are small and can be estimated analytically, at least
statistically.  The {\it HIPPARCOS} mission pioneered a second, radically
different approach.  {\it HIPPARCOS} measured the angles between of all
pairs of stars separated by about a radian.  Because each star's position
was measured against many others (which lay in sections of the sky where the
parallactic motion is substantially different), it was then possible
to determine each star's motion individually (up to a global rotation)
from a global solution of these measurements.  The global rotation was
then measured relative to the inertial quasar frame by making use of radio
stars whose positions were known in both systems.
The {\it FAME} and {\it GAIA} missions also plan to employ this method
except that the quasar tie-in will be done directly in the optical.
The {\it Space Intereferometry Mission (SIM)} will use a third
approach, which is a hybrid of the previous two.  All {\it SIM} 
astrometry will be done relative to two special classes of stars,
``grid stars'' and ``reference stars''.

	The grid will be composed of 1000-3000 astrometrically stable stars
spread over the whole sky.  By repeatedly measuring the relative
separations of these stars over the $T_m=5\,$yr span of the mission
to $\sim 10\,\mu$as precision, their absolute parallaxes and relative
positions and proper motions can be determined to $\la 3\,\mu$as and
$\la 2\,\mu$as, and $\la 2\,\mu \rm as\,yr^{-1}$, respectively.  The
resulting {\it SIM} frame can be pinned to the extragalactic inertial
frame by observing a few quasars (NASA 1999).  Absolute parallaxes, proper 
motions, and positions of all other objects can then be determined by measuring
their positions relative to the grid.  For this to work, grid stars
must be astrometrically stable on the scale of a few $\mu$as: if too
many of them prove unstable and ``drop out'' of the grid, the global
grid solution could be undermined.  
D.\ Fischer (1998 private communication)
originally argued that metal-poor K giants at $D\ga 1\,\kpc$ would make 
the only truly suitable 
grid stars because their great distance would reduce astrometric
perturbations due to unseen companions.  This view has now come to prevail.
S.\ Majewski and collaborators are conducting a photometric search for
grid candidates (Patterson et al.\ 1999).  These will ultimately be 
screened for companions using a radial velocity (RV) survey.  

	The key problem is to determine in advance
what intensity of screening is required to produce a grid that will either
have very few drop-outs, or that is structured to be able to ``paper over''
whatever dropouts it does suffer.  This problem has to date been addressed
primarily by Monte Carlo simulations (Frink et al.\ 2001; Jacobs 2000; 
Peterson 2001).

	Planet searches with {\it SIM} will be carried out by measuring
target-star positions relative to nearby ($\la 1^\circ$) astrometrically 
stable ($<1\,\mu\rm as$) reference stars.  The requirements for reference stars
are {\it qualitatively} similar to those of grid stars, and consequently I will
treat the two in parallel.  {\it Quantitatively}, however, the reference stars
are much more demanding.  First, the individual measurement error relative
to reference stars is $1\,\mu$as, an order of magnitude more precise than the
individual grid measurements.  Hence the stars must be correspondingly more 
astrometrically stable.  Second, reference stars 
must be brighter to allow for higher
precision measurements with a comparable exposure time:  Grid stars can
be $V\sim 12$, but reference stars should be 1 or preferably 2 mag brighter.
In principle, closer (and hence brighter) stars could be chosen, but, as
mentioned above, closer stars are astrometrically less stable.  
Third, reference stars 
are required at higher density.  I will argue in \S\ 5.2 that of order 8
reference star candidates must be found within $\sim 1\,\rm deg^2$ of the
target star, whereas the density of grid stars is $\la 10^{-1}\rm deg^{-2}$.
Thus, the selection of reference stars will likely include objects that are
neither metal-poor nor exceptionally 
luminous, since the density of such rare stars is simply
too low.
Finally, as I show in \S\ 5.1, grid stars that are
accelerating approximately uniformly can perform their grid functions
perfectly well.  This is not true of reference stars: if their accelerations
are at all measurable then they degrade the sensitivity to outlying planets
that are themselves detectable only due to the uniform acceleration that
they induce on the targets.  Thus, reference-star requirements for astrometric
stability are even more demanding compared to grid stars than the
relative precisions of measurement would seem to imply.  Although the
discussion here will be focused on {\it SIM}, I note that similar
considerations apply to the choice of reference stars for planet searches
using the Keck Interferometer.

	Systematic study of the reference-star problem is less advanced
than that of the grid.  Marcy (2001) and Shao (2001) discuss methods of
selecting reference stars and give initial estimates of contamination
levels.
	
	Here I present an analytic method for determining contamination
of grid stars and reference stars by undetected stellar companions.
This approach is complementary to that of the Monte Carlo studies 
mentioned above:  Monte Carlo methods can model arbitrarily complex
distribution functions in arbitrary detail, and so can potentially
capture everything that is known about a problem.  In the present
case, this intrinsic complexity seems formidable because binary
orbits are quite varied and are described by seven parameters.
However, the analytic approach has its own advantages. Primarily it
allows one to see explicitly how the results depend on the assumptions
and on the precision of the measurements.  
Thus, analytic modeling can be particularly valuable as a guide while
selection criteria and methodolgy are evolving.

	Although the underlying parameter space is quite complicated,
I show that the contamination rates can be quickly evaluated 
using a few simple formulae.  In their simplest version, these formulae
depend only on the period distribution of the companions and on their
mean mass.  This version assumes circular orbits, but I also evaluate
explicitly the correction due to eccentricities, which is small.

	For the most part, only stellar companions are considered in
this paper.  Brown dwarfs are not a practical concern because
of the ``brown dwarf desert'', the observed lack of brown-dwarf companions
to G dwarfs (Halbwachs et al.\ 2000; Marcy \& Butler 2000),
the progenitors of K giants.  Planets are not a major concern for the
grid because metal-poor giants are not expected to have many, and
because the demand for astrometric stability is not so severe.
As I briefly discuss in \S\ 5.2, planets are a significant concern for
the reference stars.  An analytic treatment of planet contamination
is feasible.  It would be broadly analogous to the one given here for 
stellar contamination.  However, the data on planetary companions do not
at present suffice to use these methods to make reliable predictions
of contamination rates.  I therefore make reference to such contamination
only for purposes of illustration and defer presentation of a detailed
treatment.

	In \S\ 2, I recapitulate some well-known statistical
results that are required for the analysis.
In \S\ 3, I analyze screening of a sample using an RV survey.
This analysis applies equally to grid stars and reference stars.  In
the next two sections on astrometric contamination (\S\ 4) and
implications (\S\ 5), I treat these two classes in parallel.  
In \S\ 6, I test the analytic formulae derived here by ``predicting''
the results of a Monte Carlo simulation by Frink et al.\ (2001).  I
find excellent agreement.  Finally, in \S\ 7, I summarize the main
formulae derived in the paper.

\section{Statistics Prelude}

	Let $F(t)=\sum_{i=1}^n a_i f_i(t)$ be a linear combination of
$n$ trial functions $f_i(t)$, each with coefficient $a_i$.  If the
parameters $a_i$ are
fit to measurements at times $t_k$, $(k=1, \ldots ,N)$, with errors $\sigma_k$,
then the covariance matrix $c_{ij}= {\rm cov}(a_i,a_j)$ is given by 
(Press et al.\ 1992)
\begin{equation}
c=b^{-1},\qquad b_{ij} = \sum_{k=1}^N {f_i(t_k)f_j(t_k)\over \sigma_k^2}.  
\label{eqn:bij}
\end{equation}
For the special case of a two-parameter fit $F(t) = a_1 + a_2 t$, where the
errors are all equal, $\sigma_k=\sigma$, equation (\ref{eqn:bij})
implies that the variance of the slope is given by,
\begin{equation}
{\rm var}(a_2)\equiv c_{22} = {\sigma^2\over N{\rm var}(t)},
\label{eqn:vara2}
\end{equation}
where ${\rm var}(t)\equiv \langle t_k^2\rangle - \langle t_k\rangle^2$.
Finally, if there are a large number of measurements uniformly spaced
over an interval $[-T/2,T/2]$, also with equal errors but now allowing
for an arbitrary number of parameters, then equation (\ref{eqn:bij})
becomes
\begin{equation}
b_{ij} = {N\over \sigma^2 T}\int_{-T/2}^{T/2} d t f_i(t)f_j(t).
\label{eqn:bijcont}
\end{equation}
For a polynomial, $F(t) = \sum_{i=1}^n a_i t^{i-1}/(i-1)!$,
these components are,
\begin{equation}
b_{ij} = {N\over \sigma^2}\,
{(T/2)^{i+j - 2}\over (i+j-1)(i-1)!(j-1)!}\qquad
(i+j\ \rm even),
\label{eqn:bijeval}
\end{equation}
and $b_{ij}=0$ otherwise.  Hence, even for a cubic 
$(n=4)$, $b$ decomposes into two $(2\times 2)$ matrices that are easily
inverted by hand.


\section{Radial Velocity Screening}

	I first ask the question: what fraction of an initial sample of
candidate grid stars or reference stars will be rejected by a RV survey 
of $N$ measurements equally spaced over a time $T_{\rm rv}$, each with a 
precision
$\sigma_{\rm rv}$.  I assume that the candidates have mass $M=1\,M_\odot$, 
and focus initially on stellar companions of mass $m$, i.e.,
$0.1\,M_\odot < m < 1\,M_\odot$.  Note that for parameters of interest,
$T_{\rm rv}\sim 5\,$yr, $\sigma_{\rm rv}\sim 60\,\rm m\,s^{-1}$, 
essentially
all companions with periods $P\la 2 T_{\rm rv}$ will easily be detected.
For example, the velocity semi-amplitude induced by an $m=0.1\,M_\odot$
companion in a $P=2T_{\rm rv}=10\,$yr orbit is $1.5\,\kms$, so that only
extremely face-on orbits, $\sin i<0.08$ would escape detection.  Since
these constitute about 0.3\% of all orientations (and so $\sim 0.02\%$
of all candidates), they will be of no interest here.

	Instead, I will work in the limit of uniform accelerations,
the magnitude of whose radial component is given by,
\begin{equation}
a_r = {G m\over r^2}\,|\cos\theta|
={63\,\rm m/s\over yr}\,\biggl({m\over 0.3\,M_\odot}\biggr)
\biggl({r\over 30\,\rm AU}\biggr)^{-2}
|\cos\theta|
\label{eqn:arad}
\end{equation}
where $r$ is the instantaneous separation of the companion and
$\theta$ is the angle between the separation vector and the line of
sight.  

	On the other hand, from equation (\ref{eqn:vara2}),
[and using $\sum_{k=1}^N k=N(N+1)/2$, 
$\sum_{k=1}^N k^2 =N(N+1)(2N+1)/6\Rightarrow {\rm var}(k)=(N^2-1)/12$],
the RV survey can measure such accelerations with precision
\begin{equation}
\sigma_a = \sqrt{12 (N-1)\over N(N+1)}{\sigma_{\rm rv}\over T_{\rm rv}}
= {16\,\rm m/s\over yr}\,
\biggl({\sigma_{\rm rv}\over 60\,\rm m/s}\biggr)
\biggl({T_{\rm rv}\over 5\,\rm yr}\biggr)^{-1}
\biggl[{N(N+1)/(N-1)\over 6.67}\biggr]^{-1/2}.
\label{eqn:sigmaa}
\end{equation}
Here $\sigma_{\rm rv}$ includes contributions from both measurement
error and intrinsic velocity variability of the stellar atmosphere, and
the expression is normalized to $N=4$ observations.
I will assume that candidates will be rejected if the observed
accelerations exceed a threshold $a_{\rm thr}=2\sigma_a$: if the
threshold were set much lower, too many candidates would be eliminated by
statistical noise.  For an ensemble of systems, $|\cos\theta|$ is
uniformly distributed over [0,1].  Hence, the fraction $f_p$ that will
pass the RV cut is
\begin{equation}
f_p(r,m) = \min\biggl\{{a_{\rm thr}\over G m/r^2},1\biggr\} =
\min\biggl\{{2 r^2\sigma_a\over G m},1\biggr\}.
\label{eqn:fpass}
\end{equation}
An ensemble of companions with semi-major axis $A$ and eccentricity
$e$, will be found equally distributed in all phases of their orbits.
Hence the fraction passing the RV screening is
\begin{equation}
f_p(e,A,m) = 1\qquad \biggl({G m\over [A(1-e)]^2} \leq a_{\rm thr}\biggr)
\label{eqn:fpeamone}
\end{equation}
\begin{equation}
f_p(e,A,m) = \biggl(1 + {3\over 2}e^2\biggr){A^2 a_{\rm thr}\over G m}
\qquad \biggl({G m\over [A(1+e)]^2} \geq a_{\rm thr}\biggr).
\label{eqn:fpeamtwo}
\end{equation}
In the intervening range, $f_p(e,A,m)$ can be written in closed form, 
but the expression is not illuminating. For the moment, I use the approximation
of circular orbits. I will address the question of eccentric orbits in
\S\ 3.2.

\subsection{Circular Orbits}

The main progenitors of K giants are G dwarfs, whose companion distribution
has been studied by Duquennoy \& Mayor (1991, DM91).  They find that
about 60\% of G dwarfs have companions, and that the log periods of these are 
roughly Gaussian distributed with $\langle \log (P/\rm days)\rangle=4.85$
and \{var$[\log (P/\rm days)]\}^{1/2}=2.3$. The companion mass distribution 
is roughly flat over the interval $0.1\,M_\odot < m < 1\,M_\odot$.
By comparison with this broad distribution over roughly 7 decades in
semi-major axis, equations (\ref{eqn:fpeamone}) and (\ref{eqn:fpeamtwo})
show that $f_p$ rises from nearly 0 to 1 in less than a decade.  Therefore,
it is appropriate to evaluate $f_p(m)$ over all $A$ as being equal to the
fraction of candidates with $A>A(m)$ where $f_p[A(m),m]=0.5$.
That is,
\begin{equation}
A(m) = \sqrt{Gm\over 2 a_{\rm thr}}
=30\,{\rm AU}
\biggl({m\over 0.3\,M_\odot}\biggr)^{1/2}
\biggl({a_{\rm thr}\over 32\,\rm m/s/yr}\biggr)^{-1/2}.
\label{eqn:aofm}
\end{equation}
For masses of $m=0.1\,M_\odot$ and $1\,M_\odot$, these values correspond
to periods of $P=10^{4.4}\,$days and $P=10^{5.0}\,$days.  That is, the
entire mass range corresponds to only about half a decade of 
the DM91 binary distribution function.
Hence, to evaluate the rejected fraction $f_{\rm rej,rv}$ over the entire
mass range, it is appropriate simply to adopt the period corresponding to
the average mass, $\langle m\rangle\sim 0.3\,M_\odot$, which is equivalent
to assuming that the DM91 cumulative distribution function is a straight
line within a bin.  The error so induced is less than 1\%, much less than
the Poisson noise in the measurement of the distribution function itself.
This yields,
\begin{equation}
f_{\rm rej,rv} = \int_0^{P_*} d P {d f_b\over d P},\qquad 
P_*^2 \equiv {4\pi^2[A(\langle m\rangle)]^3 \over G(M_\odot 
+ \langle m\rangle)},
\label{eqn:frej}
\end{equation}
where $d f_b/d P$ is the binary-period distribution function.
For the fiducial parameters I have been considering,
$P_*=10^{4.7}\,$days. Thus, substituting the period distribution of DM91
into equation (\ref{eqn:frej}), and using these fiducial parameters, 
$f_{\rm rej,rv}= 29\%$ of
the candidates would be rejected.   In the scenario I have laid out here,
another 5\% of the remaining 71\% would be rejected because statistical 
fluctuations would cause the stars to appear to accelerate at the $2\,\sigma$ 
level even when there was no real acceleration.
\subsection{Eccentric Orbits}

	To include eccentric orbits exactly would be complicated
because the transition from equation (\ref{eqn:fpeamone}) to 
(\ref{eqn:fpeamtwo})
is complicated.  However, the order of the effect can be assessed by 
noting that in both these limiting regimes, the change is accounted for
by $a_{\rm thr}\rightarrow [1 + (3/2)e^2]a_{\rm thr}$, and therefore 
making the following substitution in equation (\ref{eqn:aofm}):
\begin{equation}
a_{\rm thr}\rightarrow g_e a_{\rm thr},\qquad 
g_e\equiv 1 + {3\over 2}\langle e^2\rangle .
\label{eqn:eccone}
\end{equation}
Then, following through the remaining logic in \S\ 3.1, this leads to 
a change in the estimate of $f_{\rm rej,rv}$
\begin{equation}
\Delta f_{\rm rej,rv} = - {3\over 4}\log g_e{d f_b\over d\log P}
\bigg|_{P_*}=
-2\%\,{\log g_e\over \log(7/4)}\,{(d f_b/d\log P)|_{P_*}\over 0.11},
\label{eqn:deltafrej}
\end{equation}
where the evaluation has again been made using $f_b$ from DM91.
Because this change is so small, I will generally assume circular orbits,
but will include a 2\% adjustment for eccentricity.

\section{Fraction of Astrometric Accelerators}
\subsection{Grid Stars}

	Of the candidates that pass the RV screening and go on to
become grid stars, what fraction will have accelerations that are
detectable astrometrically? I choose a $3\sigma$ criterion to avoid 
excessive rejection due to noise.  The
astrometric acceleration $\alpha$ is given by 
\begin{equation}
\alpha = {G m\over D r^2}\,\sin\theta =
{13\,\mu{\rm as}\over\rm yr^2}
\biggl({m\over 0.3\,M_\odot}\biggr)
\biggl({r\over 30\,\rm AU}\biggr)^{-2}
\biggl({D\over 1\,\kpc}\biggr)^{-1}
\sin\theta
\label{eqn:aast}
\end{equation}
where $D$ is the distance to the grid star and all the remaining quantities
are the same as in equation (\ref{eqn:arad}).  Suppose that the astrometric
data in one direction are fit to the form
$\psi(t) = \psi_0 + \mu_0 t + (1/2)\alpha t^2 + \kappa\Pi
\cos[2\pi(t-t_0)/\rm yr]$, where
$\psi_0$ and $\mu_0$ are the position and proper motion at mid-mission,
$\Pi$ is the parallax, and $\kappa$ and $t_0$ give the phase and projection
factor of the parallax ellipse.  
Since the grid will be surveyed many (e.g., $N\sim 23$) times, the covariances
of these four parameters can be evaluated using equation (\ref{eqn:bijcont}).
First I note since the
parallax term is cyclic, whereas all the others are secular,
$b_{i4}/\sqrt{b_{ii}b_{44}}\ll 1$.  Thus, the parallax terms decouple and
so can be ignored for present purposes.  The remaining terms are
then given by equation (\ref{eqn:bijeval}), and 
the resulting covariance matrix is therefore,
\begin{equation}
c = \sigma_0^2\left(\matrix{{9\over 4} & 0 & -{30\over T_m^2}\cr
0 & {12\over T_m^2} & 0 \cr
-{30\over T_m^2} & 0 & {720\over T_m^4}\cr}\right).
\label{eqn:cij}
\end{equation}
where $T_m$ is the mission duration and $\sigma_0=\sigma/\sqrt{N}$ 
is the position error for the case where the data are fit to
uniform motion.  Hence, the acceleration error is
\begin{equation}
\sigma_\alpha = {2\,\mu{\rm as}\over\rm yr^2}
\biggl({\sigma_0\over 2\,\mu\rm as}\biggr)
\biggl({T_m\over 5\,\rm yr}\biggr)^{-2}.
\label{eqn:asterr}
\end{equation}
As stated above,
the threshold of detection is $\alpha_{\rm thr}=3\sigma_\alpha$.
We can now define a figure of merit $K$ for the relative sensitivities
of the astrometric and RV surveys
\begin{equation}
K \equiv {a_{\rm thr}\over D\alpha_{\rm thr}}=
1.13\biggl({a_{\rm thr}\over 32\,\rm m/s/yr}\biggr) 
\biggl({\alpha_{\rm thr}\over 6\,\mu\rm as/yr^{2}}\biggr)^{-1}
\biggl({D\over\kpc}\biggr)^{-1}.
\label{eqn:Kdef}
\end{equation}
Roughly speaking, this means that the astrometric survey can detect
$K=1.1$ times smaller accelerations than the RV survey and so can
detect companions at $K^{1/2}$ greater separations or $K^{3/4}$ greater
periods.  Naively, this would appear to mean that the fraction of grid
stars with detectable accelerations (relative to the pre-RV sample) is given by
\begin{equation}
f_{\rm acc} = {3\over 4}\,
{d f_b\over d \log P}\bigg|_{P_\dagger}\,{G(K)}
\label{eqn:facc}
\end{equation}
where $P_\dagger\sim P_*$ and
\begin{equation}
G(K)\rightarrow G_0(K) = \log K \qquad \rm (naive).
\label{eqn:gofk}
\end{equation}
However, this simple treatment ignores the different $\theta$-dependence
in equations (\ref{eqn:arad}) and (\ref{eqn:aast}): the astrometric
survey is sensitive to 2 components of acceleration, and corresponding
to this, the mean value of $\sin^2\theta$ is twice as high as $\cos^2\theta$
over a sphere.  In addition, the stars with low $\cos\theta$, which
are most likely to evade the RV surveillance, are the most easily detected
astrometrically.  A more rigorous treatment yields,
\begin{equation}
f_{\rm acc}(m) = \int d\log r {d f_b\over 
d \log r}\,\min\{y,\sqrt{1-(y/K)^2}\},\qquad
y(r;m)\equiv {a_{\rm thr} r^2\over G m} .
\label{eqn:facctwo}
\end{equation}
Because the second factor differs significantly from zero over only a 
relatively narrow range, the first factor can be pulled out of the integral,
which then yields equation (\ref{eqn:facc}) with,
\begin{equation}
G(K) = \int d\log y\, \min\{y,\sqrt{1-(y/K)^2}\}
= \log\bigl(K+\sqrt{K^2 + 1} \bigr),
\label{eqn:gofktwo}
\end{equation}
and with $P_\dagger$ evaluated midway between the astrometric and
RV sensitivities,
\begin{equation}
P_\dagger=K^{3/8}P_*= {2\pi K^{3/8}\over\sqrt{G(M_\odot + m)}}\,
\biggl({Gm\over 2 a_{\rm thr}}\biggr)^{3/4}.
\label{eqn:gofkthree}
\end{equation}
Again, $\log P_\dagger$ varies by only 0.6 over the whole mass range, so it
is appropriate to evaluate $f_{\rm acc}$ at $\langle m\rangle=0.3\,M_\odot$.
In this case, the correction for eccentric orbits is completely negligible
because the effect is simply to slightly displace $P_\dagger$.  However, for
the actual parameters of interest $P_\dagger$ is near the peak of the DM91
distribution (where $d f_b/d\log P=18/164\sim 11\%$), 
so the evaluation of equation (\ref{eqn:facc}) does not depend on the
exact choice of $P_\dagger$.  
For the fiducial parameters I have been using,  
$f_{\rm acc}=3.5\%$.  Hence, a fraction $f_{\rm acc}/(1-f_{\rm rej,rv})
\sim 5\%$ of the grid stars that survive RV surveillance will have
detectable astrometric accelerations.  Here, I have used
$f_{\rm rej,rv}=30\%$ to account both for eccentricity and for candidates
eliminated by statistical fluctuations.

\subsection{Reference Stars}

	Much of the foregoing derivation carries through for reference stars.
There are three major differences.  First, as I discuss in \S\ 5.2, 
reference stars are likely to be closer: I adopt $D=600\,$pc.  Second,
the reference stars will be
subject to measurements of much higher precision.  Typically, target stars
are expected to be observed relative to four reference stars at 
$N\sim 25$ epochs, each time to a total positional precision 
$\sigma_{\rm pos}\sim 1\,\mu$as.  Since each reference star is 
observed only 1/4 the time, and since their accelerations must be detected 
against each other, this implies that $\sigma_0$ in equations
(\ref{eqn:cij})  and (\ref{eqn:asterr}) should be evaluated,
\begin{equation}
\sigma_0 ={\sqrt{8/3}\sigma_{\rm pos}\over \sqrt{N}} = 0.33\,\mu{\rm as}
\biggl({\sigma_{\rm pos}\over 1\,\mu\rm as}\biggr)
\biggl({N\over 25}\biggr)^{-1/2},
\label{eqn:sigma0ref}
\end{equation}
a factor 6 times smaller than for grid stars.  
Consequently, $K$ is a factor $6/0.6$ times larger,
$K=11$.

	Third, depending on the ultimate design of SIM, the reference stars 
may be measured relative to the target only along one dimension (`ParaSIM')
or in two dimensions (`Shared Baseline').  In the latter case, the analysis
is identical to that given in \S\ 4.1.  Using $K=11$ and equations
(\ref{eqn:facc}) and (\ref{eqn:gofktwo}), I then find
$f_{\rm acc} = 11\%$, implying that a fraction $f_{\rm acc}/(1-f_{\rm rej,rv})
\sim 16\%$ of reference stars that survive RV surveillance will have
measurable accelerations.

For the former case (which at the present time appears less likely), 
$\sin\theta$ in equation (\ref{eqn:aast}) 
should be replaced by  $\sin\theta|\cos\phi|$ where $\phi$ is a random
angle on a circle.  In principle this means that equation (\ref{eqn:facctwo})
should be replaced by a more complicated integration.  In fact, this
is unnecessary for the case $K\gg 1$, which is of relevance here.
Since the RV and astrometric measurents are both one-dimensional, 
the fraction of stars with detectable 
RV accelerations at radius $r_{\rm rv}$ will be exactly the same as those
with detectable astrometric accelerations at 
$r_{\rm ast} = K^{1/2} r_{\rm rv}$.  Hence, if these two detection
processes could be regarded as completely independent of one another,
equations (\ref{eqn:facc}) and (\ref{eqn:gofk}) would hold.  As discussed
following equation (\ref{eqn:gofk}), these detection processes are not
generally independent.  However, for $K\gg 1$, they are approximately 
independent because at the radii where the radial acceleration
is detectable at all, there is only a very small chance that the
astrometric acceleration will be undetectable.  Applying equations
(\ref{eqn:facc}) and (\ref{eqn:gofk}), and taking $K=11$, I find
$f_{\rm acc} = 9\%$, implying that $f_{\rm acc}/(1-f_{\rm rej,rv})
\sim 12\%$.

\section{Implications}
\subsection{Grid Stars}

	In \S\ 4.1, I showed that about 5\% of SIM grid stars that 
survive RV selection will have measurable astrometric accelerations.
If it were necessary to eliminate these stars from the grid well
into the mission, then it would be necessary to build redundancy into
the grid to ``paper over'' the resulting holes.  In fact, 
this is not necessary.  The fundamental reason is
that, as I show below, while some stars surviving RV selection may have 
detectable accelerations, almost none have detectable jerks.  If this
is the case, the accelerating stars can be fit to seven parameters 
(including two
components of acceleration) instead of the usual five.  From equation
(\ref{eqn:cij}) this acceleration measurement decouples completely from the 
proper-motion measurement, and as discussed directly above equation
(\ref{eqn:cij}), it decouples from
the parallax measurement as well.  Thus, the grid parallax and 
proper-motions (the main reasons for having a grid) are not significantly 
affected by fitting for acceleration.  The error in the mean position is
increased by a factor $c_{1,1}^{1/2}/\sigma_0= 1.5$. 
However, this error is more than an order
of magnitude below any known requirement.  For example, the positions
are used for the tie-in to the radio reference frame, but the latter is
known only to $20\,\mu$as, and the individual positions of quasars are
much more poorly determined.  Moreover, this slight degradation in
positional error affects only 5\% of the grid stars.

	What fraction of grid stars surviving RV selection will have
detectable jerks?  Noting that the parallax measurement again decouples
(see \S\ 4.1),
I write the remaining positional dependence in one direction as
$\psi(t) = \psi_0 + \mu_0 t + (1/2)\alpha_0 t^2 + (1/6)j t^3$.  I then evaluate
$b_{ij}$ using equation (\ref{eqn:bijcont}) and invert the resulting
$(2\times 2)$ (proper-motion,jerk) submatrix, thus finding
\begin{equation}
\sigma_j = \sqrt{100800}{\sigma_0\over T^3_m}=
{5\,\mu \rm as\over yr^3}
\biggl({\sigma_0\over 2\,\mu\rm as}\biggr)
\biggl({T_m\over5\,\rm yr}\biggr)^{-3}.
\label{eqn:sigmaj}
\end{equation}
For circular orbits of period P, the astrometric jerk is given by
\begin{equation}
j = {m\over M_\odot}{r\over D}\biggl({2\pi\over P}\biggr)^{3}\simeq
{8\,\mu\rm as\over yr^3}
\biggl({m\over 0.1\,M_\odot}\biggr)
\biggl({r\over 10\,\rm AU}\biggr)^{-7/2}
\biggl({D\over\kpc}\biggr)^{-1},
\label{eqn:jerk}
\end{equation}
where to be conservative I have assumed that the jerk is in the plane
of the sky and where I have written the formula in a way that is strictly
appropriate only for $m\ll M_\odot$.  I will show below that this is the 
regime of the greatest concern.  Hence, for the fiducial parameters,
jerk is detectable at the $2\,\sigma$ level provided that
$r< 9\,{\rm AU}(m/0.1\,M_\odot)^{2/7}$.  In the uniform-acceleration
approximation, such accelerations escape RV detection only a fraction
$14\%(m/0.1\,M_\odot)^{-3/7}$ of the time [see eqs.\ 
(\ref{eqn:arad})--(\ref{eqn:fpass})].  
However, at $r=9\,$AU, the uniform-acceleration approximation
is already starting to break down and, as I discussed in \S\ 3, 
for $r<4.5\,$AU, the RV survey almost never fails to detect stellar
companions.  I therefore
estimate that for 1/2 dex in $\log P$, the RV survey misses
$\sim 10\%$ of grid candidates that go on to show marginally detectable
jerk, i.e., about 0.4\% of grid stars.

	To understand the effect of eccentricity, note that the worst
case is a companion at periastron $r$ of a highly eccentric orbit.  Then
the acceleration is the same as for a circular orbit at $r$, but the
jerk is larger by $\sqrt{2}$.  Since stars spend little time at periastron
and since the effect itself is small, I ignore it.

	Finally, I have ignored planetary companions of grid candidates.
The plan is for the grid to
be composed of metal-poor K giants, which are not expected
to have planets for several reasons.  First, planet frequency 
among G stars (the progenitors of K giants) in the solar neighborhood
is highly correlated with metallicity (Gonzalez 1997; 
 Gonzalez, Wallerstein \& Saar 1999).
Second, planets occur in 47 Tuc at a rate that is much lower than
that of solar-metallicity stars, and which is consistent with zero 
(Gilliland et al.\ 2000).  Third, if gas giant planets grow from
rock and ice cores, as most current theories suggest, it is difficult
to see how they would get started in a metal poor environment.
This optimistic assessment could prove wrong, but if so, it will
become evident early in the RV survey of candidates, which would
give plenty of time to modify strategy.  For now, it is reasonable
to suppose that such planets will be rare or non-existent.

	In brief, with the very reasonable fiducial parameters adopted
here, I expect that RV screening can identify grid stars whose
chance of being corrupted by an undetected companion is $\ll 1\%$.

\subsection{Reference Stars}

	As I discussed in the introduction, 
contamination is a much more severe problem for 
reference stars than it is for the grid.  First, the astrometric measurements
are substantially more precise, meaning that they are more sensitive to
contaminants.  Second, the reference stars need to be brighter (to achieve
this greater astrometric precision in a short exposure), which generally
means that they need to be closer.  This in turn increases the astrometric
contamination, which scales $\propto D^{-1}$.  Third, their required surface
density on the sky is higher, which means one cannot typically find
metal-poor halo stars to serve as reference stars.  Hence, a significant
fraction is likely
to have planets.  Finally, uniform acceleration seriously undermines
the function of reference stars whereas, as we saw in \S\ 5.1, it is
perfectly acceptable for the grid.  Here I employ the formalism developed
above to illuminate these problems.

	In order to be sensitive to very low levels of acceleration due
to distant planets, the observer must find at least two reference 
stars\footnote{In the case of ParaSIM, which measures relative offsets
in only one direction, two reference stars must be found in each of
two orientations.  I ignore explicit consideration of ParaSIM here, but
it is straight forward to extend the results presented to this case.}
that are themselves unaccelerating.  This
is because, if the target star and the reference star are found to be
accelerating relative to one another, there is no way to determine which 
is truly accelerating and which (if either) is in uniform motion.  It
is only by finding two reference stars that are not accelerating relative
to one another that one can have reasonable confidence that it is they,
and not the target, that are in uniform motion.  If two such stars cannot
be found, then one still has sensitivity to planets
with periods $P \la T_m$, but not to outlying planets, including outlying
companions of planets with short periods (i.e., planetary systems).

	The first point then, is that the RV survey must begin
with enough candidates to have good prospects (say 95\% probability)
of finding at least two candidates that have no detectable RV acceleration.
I showed in \S\ 3, that
with a 5 year, $60\,\rm m\,s^{-1}$ survey, one could expect 30\% rejection
due to stellar companions and statistical fluctuactions.  I have not
evaluated planet contamination in this paper, 
but for illustration, I will assume that an additional 7\% of
candidates are eliminated by the RV survey due to planets.  This means
that if one wants to wind up with 2 reference stars, one should begin 
with 8 candidates.

	The second point is that if these are the only reference
stars used, then the probability that one of them will prove to be
an astrometric accelerator is high.  In \S\ 4.2, I showed that 16\%
would have detectable accelerations due to stellar companions.  
However, there will be additional losses due to planets.  For example, 
a Jupiter-mass planet at 3 AU would generate a velocity semi-amplitude
of only $17\sin i\,\rm m\,s^{-1}$, well below the detection threshold of the
RV survey, but an astrometric semi-amplitude of $5\,\mu$as, quite easily
detectable.  Hence, an additional 5\% loss due to planets is quite 
plausible.  If so, the probability of at least one of the two reference
stars being an accelerator would be 38\%.

	There are basically four alternatives. 1) Accept that for more than
a third of 
the target stars there will be sensitivity to closed planetary orbits,
but not to distant planets.  2) Accept fainter reference stars to increase
$D$ and so decrease astrometric contamination, but thereby degrade the 
astrometric precision and so the sensitivity to low-mass
planets.  3) Increase the number of reference stars to increase
robustness.  4) Increase surveillance efforts to reduce contaminants.

	I have not much to say about the first two options, save that 
I would be disappointed if they were adopted.  The third can be 
implemented at fairly low cost.  If 8 candidates are initially surveyed, 
then there is an 87\% chance that at least
three will survive the RV survey, and 66\% chance that four will survive.  
With three reference
stars instead of two, the probability that at
least two will survive rises from 62\% to 89\%.  Of course, there is
then also a high additional probability (39\%) that one of these three
will fail, in which case the errors for detecting distant (but not close)
planets would increase by $\sqrt{3/2}$.  This is still a lot better
than a total loss.  The situation would be even more favorable with
four reference stars.  

	Finally, a more intensive RV study would be expensive in
terms of big-telescope time, but would be effective against both
stellar and planetary companions.  For example, reducing $\sigma_{\rm rv}$
from 60 to 20 m/s would decrease the number of stellar contaminants
by $\Delta\ln G(K)\sim 36\%$.  It would not be possible
to go below 20 m/s because this is the typical scale of photospheric
fluctuations of K giants (Frink et al.\ 2001).  However, if need be
one could achieve the same effect with multiple measurements
(assuming, as is almost certainly the case, that these fluctuations
have power overwhelmingly on short timescales so that they do not
couple to the acceleration measurements).

\section{Comparison with Monte Carlo}

	Comparison with the Monte Carlo simulation of Frink et al.\ (2001)
permits a direct check of the foregoing calculation, but first it is
necessary to translate their parameterization into the one used here.
They consider a RV survey with N=2 epochs separated by $T_{\rm rv}=5\,$yr,
each with $\sigma_{\rm rv}= \sqrt{2}\times 20\,\rm m\,s^{-1}$ (including
measurement errors and intrinsic instability of the stellar atmospheres).
Hence, from equation (\ref{eqn:sigmaa}), $\sigma_a=8\,\rm m\,s^{-1}$.
I focus on the case of $\chi^2_{\rm limit}=4$ (their Fig.\ 6) corresponding
to $a_{\rm thr}=\sqrt{8}\sigma_a$.  They demand that the rms scatter of the
astrometric measurements be less than $1\,\mu$as for a $T_m=5\,$yr mission,
which corresponds to 
$\alpha_{\rm thr} = \sqrt{720}\mu{\rm as}/T_m^2 = 1.07\,\mu\rm as\,yr^{-2}$,
and place their grid stars at $D=2\,\kpc$ (S.\ Frink, private communication
2001).  From equation (\ref{eqn:Kdef}) these values imply $K=2.23$,
hence $G(K)=0.67$.  From equation (\ref{eqn:gofkthree}), I find,
$\log P_\dagger =5.2$, which I then apply to their adopted binary distribution 
function $df_b/d\log P = 0.087\exp[-(\log P -4.8)^2/2\times 2.3^2]$, to
find $df_b/d\log P_\dagger=0.086$. Equation (\ref{eqn:facc}) then predicts that
4.3\% of their original sample should be found to be accelerators, or
since 25\% of that sample was rejected by their RV selection, 
$4.3\%/0.75=5.8\%$ of the grid stars.  This compares with the value 
$5.5\pm 0.5\%$ shown in their Figure 6, i.e., excellent agreement.

\section{Summary of Formulae}

	I find that the fraction of grid-star or reference-star candidates
that are eliminated by RV surveillance
is equal to the fraction with binary companions having $P<P_*$,
where $P_*$ is given by equations (\ref{eqn:aofm}) and (\ref{eqn:frej}) 
in terms of mean mass of the companion distribution and the threshold of
acceleration detection, $a_{\rm thr}={\cal N}\sigma_a$.  Here $\sigma_a$
is the acceleration-measurement error, given by equation
(\ref{eqn:sigmaa}).  (I used ${\cal N}=2$.)\ \ 
There is a small correction for eccentricity given explicitly by equation 
(\ref{eqn:deltafrej}), and of course some candidates will be falsely
rejected due to statistical fluctuations, depending on the choice of
${\cal N}$.

	The fraction of initial candidates that survive RV surveillance but 
nevertheless have detectable astrometric accelerations is given by equation
(\ref{eqn:facc}), $f_{\rm acc}= (3/4)(d f_b/d\log P_\dagger)G(K)$.  The
first term is simply the differential binary distribution evaluated
at $P_\dagger$, which latter
is given explicitly by equation (\ref{eqn:gofkthree}).
The second term is given by equation (\ref{eqn:gofktwo}),
$G(K) = \log(K+\sqrt{K^2+1})$ as a function of $K$, which characterizes
the ratio of RV to astrometric sensitivies and is given by (\ref{eqn:Kdef}).
In this case the correction for eccentricity is negligible.

	I find that a relatively modest RV survey can remove all but
$\sim 5\%$ of grid-star candidates that will go on to show detectable
accelerations, and all but $\sim 0.4\%$ of those that will show
detectable jerks.  I argue that it is only the latter very small
fraction that must be eliminated from the grid.

	For reasons summarized in \S\ 5.2, reference-star selection is
much more demanding than grid-star selection.  I find that to have a
95\% chance of ultimately locating 2 reference stars 
(the minimum required to be sensitive to distant planetary companions
of the target star) 8 candidates must be initially surveyed for each 
target star.

\begin{acknowledgements}
I thank J.\ Catanzarite, D.\ DePoy, S.\ Frink, S.\ Majewski, G.\ Marcy, 
A.\ Quirrenbach, M.\ Shao, R.\ Swartz, and S.\ Unwin for 
valuable discussions.
This work was supported by NSF grant AST~97-27520, by JPL
contract 1226901, and by
a grant from Le Centre Fran\c cais pour L'Accueil et Les \'Echanges
Internationaux.

\end{acknowledgements}

\end{document}